\documentclass[aj_pt4]{aastex}
\usepackage{amsmath}
\usepackage{booktabs}
\usepackage{threeparttable}
\def\etal {et al.~}
\def\eg   {e.g.,~}
\newbox\grsign \setbox\grsign=\hbox{$>$} \newdimen\grdimen \grdimen=\ht\grsign
\newbox\laxbox \newbox\gaxbox
\setbox\gaxbox=\hbox{\raise.5ex\hbox{$>$}\llap
     {\lower.5ex\hbox{$\sim$}}}\ht1=\grdimen\dp1=0pt
\setbox\laxbox=\hbox{\raise.5ex\hbox{$<$}\llap
     {\lower.5ex\hbox{$\sim$}}}\ht2=\grdimen\dp2=0pt

\shorttitle{Gemini molecular cloud}
\shortauthors{Yingjie \etal}

\newcommand{\co}{$^{12}$CO }                             
\newcommand{\xco}{$^{13}$CO }                            
\newcommand{\xxco}{C$^{18}$O }                           

\begin{document}

\title{CO Core Candidates in the Gemini Molecular Cloud }

\author{Yingjie Li \altaffilmark{1, 2, 3},   Ye Xu \altaffilmark{2},  Ji Yang \altaffilmark{2}, Xin-Yu Du  \altaffilmark{2, 3}, Deng-Rong Lu \altaffilmark{2}, Fa-Cheng Li \altaffilmark{2, 3}}

\altaffiltext{1}{Qinhai Normal University, China; Xining 810000, liyj@pmo.ac.cn}
\altaffiltext{2}{Purple Mountain Observatory, Chinese Academy of Sciences, China; Nanjing 210008, xuye@pmo.ac.cn}
\altaffiltext{3}{Graduate University of the Chinese Academy of Sciences, 19A Yuquan Road, Shijingshan District, Beijing 100049, China}

\begin{abstract}
We present observations of a 4 squared degree area toward the Gemini cloud obtained using J = 1-0 transitions of $^{12}$CO, $^{13}$CO and C$^{18}$O. No C$^{18}$O emission was detected. This region is composed of 36 core candidates of $^{13}$CO. These core candidates have a characteristic diameter of 0.25 pc, excitation temperatures of 7.9 K, line width of 0.54 km s$^{-1}$ and a mean mass of 1.4 M$_{\sun}$. They are likely to be starless core candidates, or transient structures, which probably disperse after $\sim$10$^6$ yr.
 \end{abstract}


\section{Introduction}
Over the past several decades, relative to larger molecular gas surveys in Galactic plane, very few similar systemic studies were towards high galactic latitudes. The Columbia survey conducted a large area survey up to a latitude \emph{b} of $\pm$ 35$\degr$ \citep{TDH2001},  however its relative poor angular resolution ($\sim 8\arcmin$) is much larger than a typical core.

The study of high-latitude molecular gas was active in the 1980s and 1990s, and had made some achievements: firstly, the high-latitude molecular gas was researched by various of molecules, such as CO and its isotopes, NH$_3$, H$_2$CO \citep{MBW1988, T1993a, T1993b}, etc.; secondly, in terms of its physical conditions, it is similar to the diffuse cloud in the galactic plane, and in terms of its chemical abundance, it is similar to the cold dark cloud in the galactic plane \citep{TRX1989}; and thirdly, the high-latitude molecular gas was believed in the vicinity of sun \citep{MBM1985}; etc.

We are carrying out a large-scale survey towards high-latitude molecular gas survey in the northern sky with a resolution of $\sim50\arcsec$. Our aims are to understand the structure, stability and physical conditions of molecular cores in the local area, between longitudes \emph{l} from -10$\degr$ to 280$\degr$ and \emph{b} over $\pm$ 5$\degr$. For a starting, here we report our observations of the Gemini molecular cloud that is centered at \emph{l} = 200$\degr$  and \emph{b} = 12\degr, and was also observed by the Columbia survey \citep{TDH2001}, and no other literature made a further effort in this region in details. When it comes to ``Gemini molecular cloud'', one usually associates it with ``Gem OB 1''. However, it has nothing to do with Gem OB 1, since its $V_{lsr}$ is much less than Gem OB 1, and it is located at the east-north of Gem OB 1 with angular distance about 12$\degr$, and the distances to those two clouds are markedly different. With a resolution of 51$\arcsec$, which is about a factor of 10 higher than the Columbia survey, we are able to distinguish a molecular core with scale of about 0.10 pc at a distance of 400 pc, which is estimated by combining several ways. See details in section 3.2.

\section{Observations and Data Reduction}

We observed \co(1-0), \xco(1-0) and \xxco(1-0) with the Purple Mountain Observatory Delingha (PMODLH) 13.7 m telescope from May 5 to June 21 and December 18 to December 30, 2014. These three lines  were simultaneously observed with the 9-beam superconducting array receiver (SSAR) working in the sideband separation mode and using the fast Fourier transform spectrometer \citep{YS2012}.

Our observations were made in 16 cells of dimensions 30$\arcmin\times$30$\arcmin$, which covered an area of 4 square degrees (248 pc$^2$ at distance of 400 pc). The cells were mapped using the on-the-fly (OTF) observation mode with the standard chopper wheel method for calibration \citep{PB}. In this mode, the telescope beam scanned along lines of galactic longitude and galactic latitude at a constant rate of 50\arcsec/sec, and the receiver records spectra every 0.3 sec. Each cell was scanned in both the galactic longitude and the galactic latitude directions to reduce the fluctuation of noise perpendicular to the scanning direction. The typical system temperature ($T_{sys}$) during observations was $\sim$250 K for \co and $\sim$160 K for \xco and C$^{18}$O. Finally, we calibrated the antenna temperature ($T^{\ast}_{A}$) to the main beam temperature ($T^{\ast}_{R}$) with a main beam efficiency ($\eta_{mb}$) of 46\% for \co and 51\% for \xco and C$^{18}$O. A summary of the observation parameters is listed in Table 1.

 \begin{table}[!ht]
 \begin{center}
 \textbf{Table 1}

 Observation Parameters

 \end{center}
 \centering
 \begin{threeparttable}
 \begin{tabular} {c c c c c c c} 
  \hline \hline
  Line &  $\nu _{0}$ & HPBW & $T_{sys}$ & $\eta_{mb}$ & $\delta v$ & $T^{\ast}_{R}$ rms noise \\
  (J=1-0) & (GHz) & (\arcsec) & (K) &  & (km s$^{-1}$) & (K) \\ 
  \hline
  \co & 115.271204 & 49$\pm$2 & 220-300 & 46\% & 0.16 & 0.28 \\
  \xco & 110.201353 & 51$\pm$2 & 140-200 & 51\% & 0.17 & 0.13 \\
  \xxco & 109.782183 & 51$\pm$2 & 140-200 & 51\% & 0.17 & 0.19 \\
  \hline
 \end{tabular}

 \end{threeparttable}
 \end{table}

\section{Result}

\subsection{General Distribution}

Figure 1 shows the distribution of \co emission in the Gemini molecular cloud. The strongest emission is located at (\emph{l}, \emph{b}) = (200.26\degr, 11.57\degr) and (\emph{l}, \emph{b}) = (200.63\degr, 11.75\degr) with integrated intensities of 10.7 and 10.5 K km s$^{-1}$, respectively. Similarly, Figure 2 presents an integrated intensity map of $^{13}$CO. The strongest emission is located at (\emph{l}, \emph{b}) = (199.55\degr, 11.88\degr) and (\emph{l}, \emph{b}) = (199.95\degr, 11.78\degr) with integrated intensities of 1.1 and 1.0 K km s$^{-1}$, respectively. The positions of the emission peak of both lines are not exactly matched. Some emission in \co has no counterpart in \xco, implying that the \co is more widely distributed and extended than $^{13}$CO. Unlike $^{12}$CO, \xco shows no continuous distribution, and only some condensations. Generally the gas distribution looks diffuse. Strong emission is only concentrated at several small locations. In the \co emission, clearly there are some filamentary structures that contain a few cores, while the \xco emission is largely composed of separate, individual core. No C$^{18}$O emission was detected in a mean rms noise level of 0.19 K. We have averaged all of the spectra together to provide a single aggregate measure of the \xxco brightness over the cloud, which shows RMS of 0.01 K, and still unable to find any \xxco emissions. The reason may be that the emission of C$^{18}$O is too faint to be detected, and the upper limit for average $T^{\ast}_{R}$ of \xxco is 0.01 K.

\begin{figure}[ht!]
\centering
\includegraphics[width=10cm,height=10cm,angle=0]{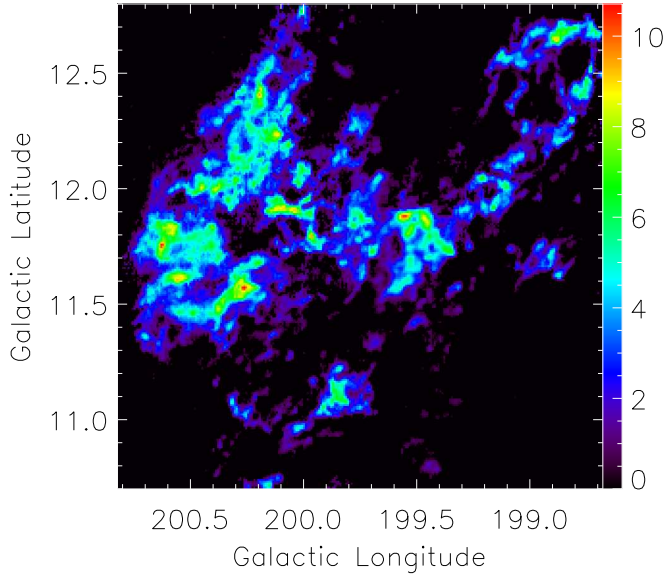}

\caption{Integrated intensity map of \co, integrated between -7 and 7 km s$^{-1}$ and scaled between 0 and 11.0 K km s$^{-1}$.}
\end{figure}

\begin{figure}[t]
\centering
\includegraphics[width=10cm,height=10cm,angle=0]{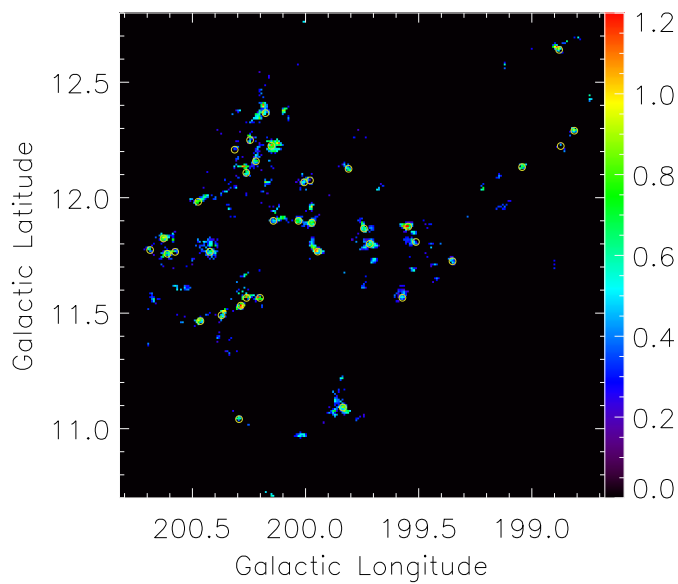}

\caption{Integrated intensity map of \xco, integrated between -3 and 5 km s$^{-1}$ and scaled between 0 and 1.2 K km s$^{-1}$. The yellow circles show the positions of core candidates in $^{13}$CO.}
\end{figure}

\begin{figure}[!ht]
\centering
\includegraphics[width=10cm,height=7cm,angle=0]{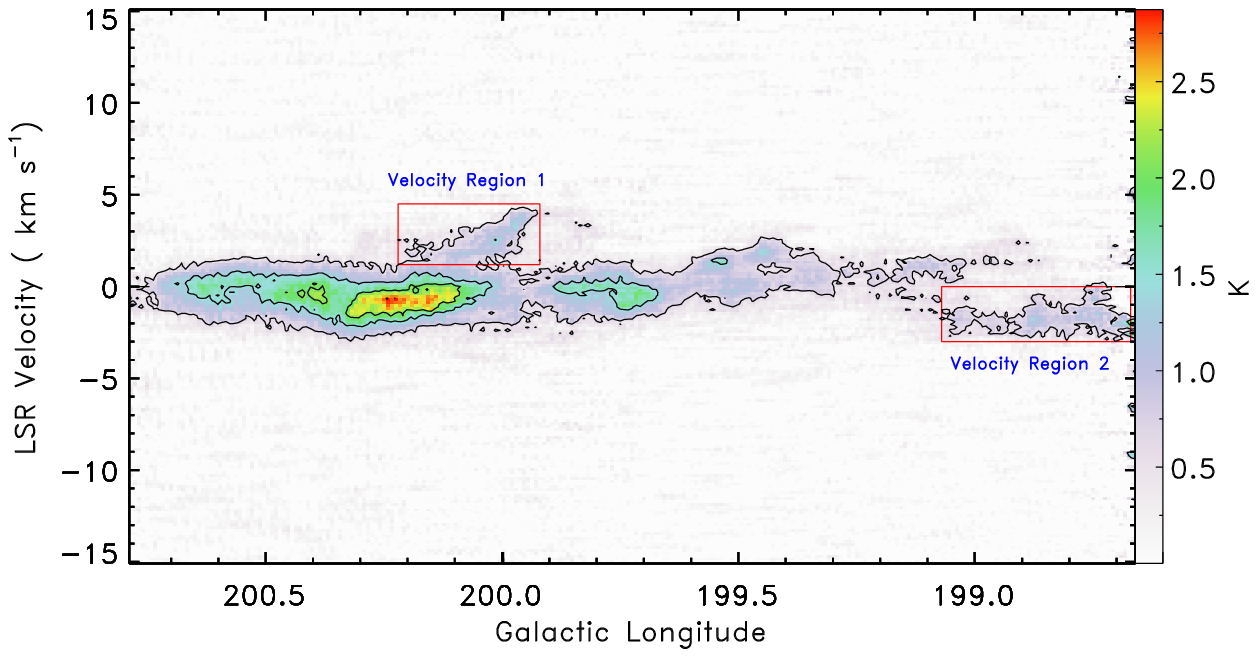}

\caption{Longitude-velocity map in \co.}
\end{figure}

Figure 3 presents the longitude-velocity map of $^{12}$CO. The longitude-velocity map of \co suggests that $V_{lsr}$ is confined to (-5, 5) km s$^{-1}$ and mainly around $\sim$ 0.0 km s$^{-1}$. A second velocity component (hereafter, velocity region 1) is centered at $\sim$ 2.5 km s$^{-1}$, ranging from $l\approx 199.9\degr$ to $l\approx 200.2\degr$, and a third one (hereafter, velocity region 2) is centered at $\sim$ -1.7 km s$^{-1}$ in the range of $l\lesssim199.1\degr$. There are large offsets from 0.0 km s$^{-1}$ in the two velocity components. In addition, we find a velocity gradient of $\sim$ 0.3 km s$^{-1}$ pc$^{-1}$ in the range of $l\approx 200.0\degr$ to $l\approx 200.7\degr$.

\begin{figure}[!ht]
\centering

\includegraphics[width=8cm,height=10.8525cm,angle=0]{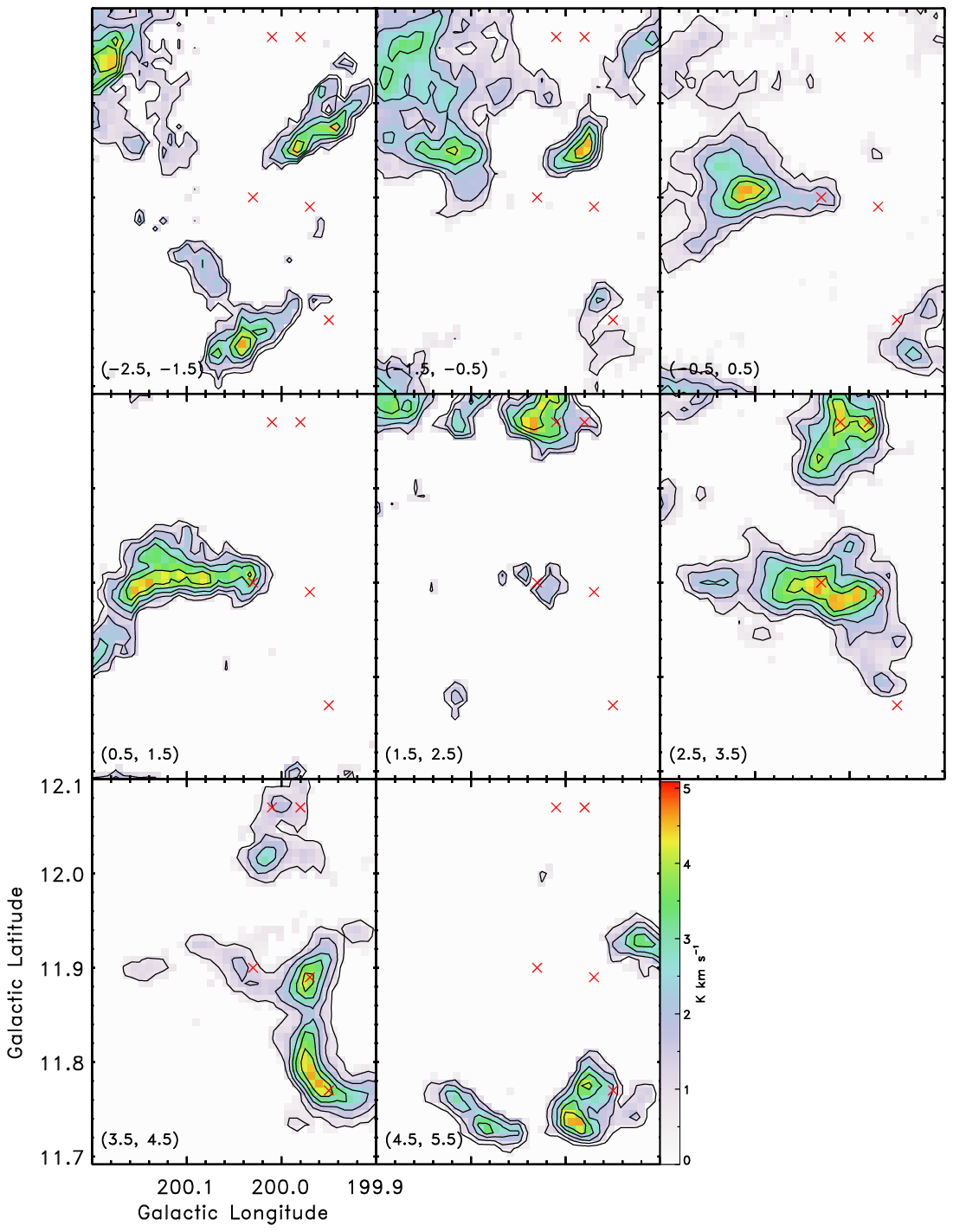}
\includegraphics[width=8cm,height=10.66667cm,angle=0]{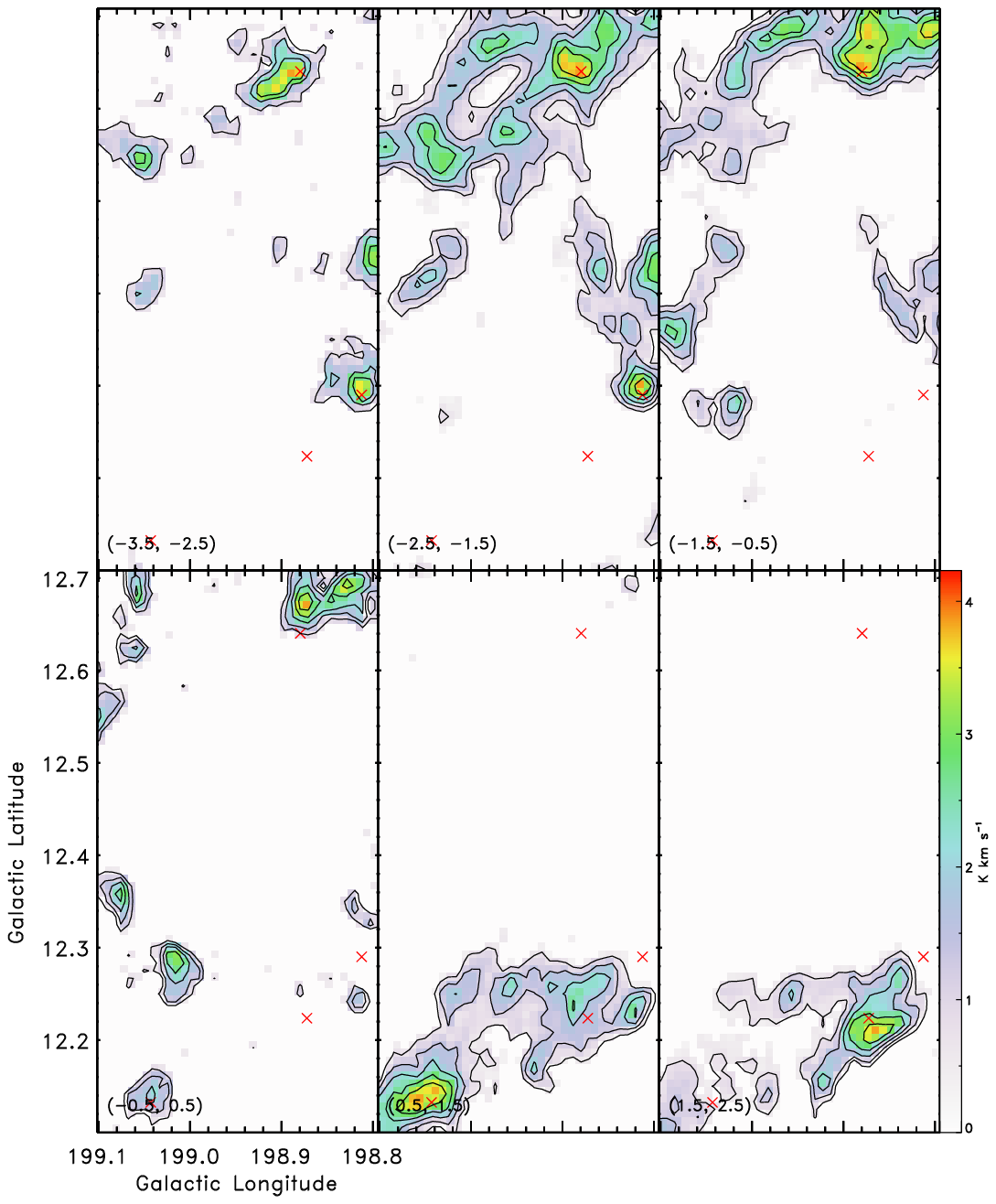}

\caption{Channel Maps of $^{12}$CO in part regions. The two regions range from $(l, b)=(199.89\degr, 11.70\degr)$ to $(200.20\degr, 12.11\degr)$ in the left map, and from $(l, b)=(198.79\degr, 12.10\degr)$ to $(199.10\degr, 12.71\degr)$ in the right map. These sections corresponding to velocity region 1 \& 2 in the longitude-velocity map. Each map was integrated over a velocity interval of 1 km s$^{-1}$, and the velocity ranges (in unit of km s$^{-1}$) of each map are marked in the lower-left corner. The lowest contour is 3$\times$RMS, and the contour interval is equal to (maxvalue of each map$-$minvalue of each map)/5. The core candidates of \xco fallen into the two regions are also indicated by red crosses. Note that the velocity range in both maps are the velocity range of the signal in each corresponding regions.}
\end{figure}

At first, to make clear of the velocity structure in those two velocity regions, the channel maps in Figure 4 illustrate the velocity structure of the molecular emission in part regions corresponding to those two velocity regions. The left maps range from $(l, b)=(199.89\degr, 11.70\degr)$ to $(200.20\degr, 12.11\degr)$, while the right ones ranges from $(l, b)=(198.79\degr, 12.10\degr)$ to $(199.10\degr, 12.71\degr)$. Both sections correspond to velocity regions 1 \& 2 in Figure 3. For ease of presentation, we refer to them as map 1 and map 2, respectively.

Map 1 shows a montage of \co emission distributions with velocities ranging from -2.5 to 5.5 km s$^{-1}$ in every 1 km s$^{-1}$, which indicates a drastic change of morphology: no cloud component appears in more than 2 successive channels. Similarly, map 2 shows the montage of \co emission with velocities ranging from -3.5 to 2.5 km s$^{-1}$ in every 1 km s$^{-1}$ which presents a velocity distribution of part of the filamentary components. This montage contains two distinct components of $V_{lsr}$ that shift from -2.5 to 2.5 km s$^{-1}$. These complicated features in the velocity distributions likely indicate that the cloud consists of many components with velocities differing by $\sim$ 5 km s$^{-1}$. Similar to map 1, map 2 also shows a big change of morphology; for instance, the bottom cloud components appear in only 2 channels. The big change of morphology in the channel maps may be triggered by stochastic processes between clouds such as collisions and chaotic magnetic fields, rather than ordered motions such as rotation. Furthermore, unlike map 1, map 2 shows some filamentary structures, especially in the velocity range of -1.5 to -0.5 km s$^{-1}$ and from 0.5 to 1.5 km s$^{-1}$. In addition, \co emissions seldom simultaneously appear at discontinuous channels in both maps.

And then, we present the channel map of the entire region of the Gemini Molecular cloud in Figure 5. It shows that most \co emissions are confined to (-2, 2) km s$^{-1}$, and indicates that the morphology in Figure 5 changes less than those two velocity regions shown in Figure 4.

\begin{figure}[!ht]
\centering

\includegraphics[width=12cm,height=16cm,angle=0]{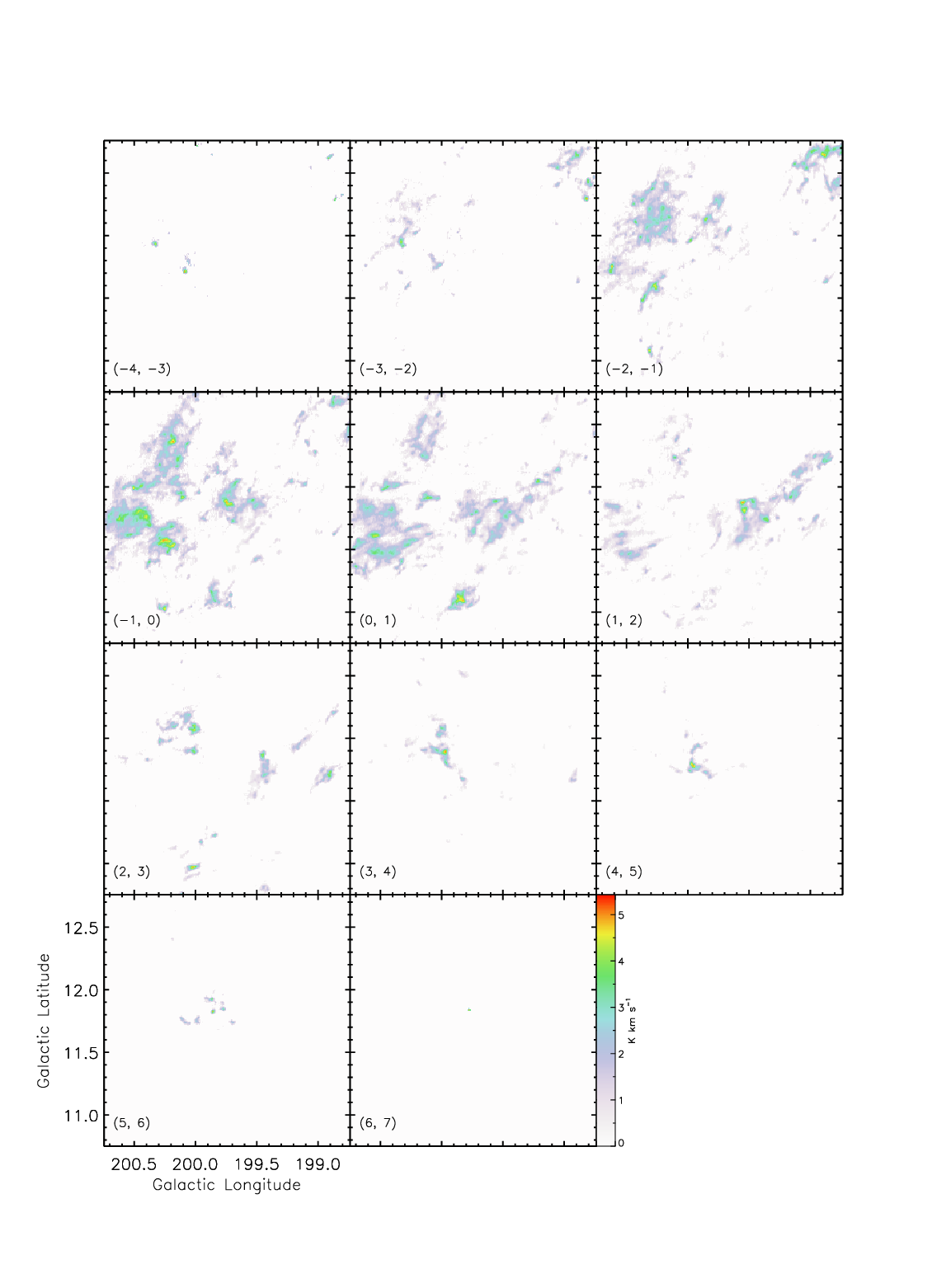}

\caption{Channel Maps of $^{12}$CO in the whole regions. Each map is integrated over a velocity interval of 1 km s$^{-1}$, and the velocity ranges (in unit of km s$^{-1}$) of each map are marked in the lower-left corner.}
\end{figure}

\subsection{Distance Estimate}

In order to obtain the radius of a core candidate, denoted by $R_{13}$ for $^{13}$CO, the mass (based on local thermodynamic equilibrium (LTE)), denoted by $M$, and the virial mass, denoted by $M_{V}$, the distance to the core candidate is indispensable. Kinematic distance is a widely used distance. Whereas the $V_{lsr}$ of the cloud is around 0 km s$^{-1}$ and cannot be effectively used to estimate it, therefore, we have tried other four methods to work it out.

Firstly, we try to find stars in the line of sight of the Gemini molecular cloud, and get a Tycho 2 star---TYC 1349-01421-1---with distance of 131 pc \citep{ARS2006}, which is located at (07:11:05.29, +17:00:45.9) in J2000 equatorial coordinate system. The $A_{K_{s}}=0.174$, and $A_{K}=0.95 A_{K_{s}}$ = 0.165, which indicates that this star is slightly reddened and may be also a foreground star as its $A_{K}<0.3$ \citep[see e.g.,][]{LAL2011}.

Secondly, for a given cloud, the expectation value of the height above the plane is 0.798 $\sigma_{z}$, where $\sigma_{z}$ is the Gaussian scale of the gas and is about 75 pc, and the distance to a cloud is 0.798 $\sigma_{z} \csc b$, where $b$ is the galactic latitude \citep[see e.g.,][]{MBM1985}. Using this method, we estimate that the distance is around 300 pc. However, this method is generally applied to clouds with $b \geq 25 \degr$. Whereas, molecular gases are mainly in galactic plane.

Thirdly, the facts that the Gemini molecular cloud is centered at $l=200\degr$ and the $V_{lsr}$ is around 0 km s$^{-1}$ indicate that this cloud could be nearby, for instance, like Taurns, Lam Ori or Califirnia giant Molecular cloud whose distances are 140 pc, 400 pc and 450 pc, respectively \citep{JLM1997, LLA2009, LMD2000}.

Lastly, we use the interstellar extinction distribution to estimate the distance to Gemini molecular cloud. We follow the approach applied by \citet{MRR2006} and use the stellar population synthesis model of the Galaxy constructed in Besan\c{c}on \citep{RRDP2003}. Figure 6 shows the result. We find three 2MASS sources, J07084534+1610175, J07125749+1647429 and J07104500+1755493, which are in line of sight of the \xco clumps of Gemini molecular cloud and have $A_{K_{s}}=0.67(J-K_{s})$ of 0.261, 0.526 and 0.532, respectively. And the distance estimated by 2MASS source J07084534+1610175 is 430$\pm$170 pc, with reliability of 0.68. However, this source has $A_{K}=0.95 A_{K_{s}}$ = 0.248, which may indicate that this source may be a foreground star as its $A_{K}<0.3$ \citep[see e.g.,][]{LAL2011}, and therefore may slightly underestimate the distance. By using other two 2MASS sources which are definitely background stars, we find the distance reach to 2.6 kpc. While, both of those two 2MASS stars are about $\gtrsim$ 1 mag dimmer than J07084534+1610175, and have poorer Signal-to-Noise ratio with a factor of $\gtrsim$ 3.2.

The Tycho 2 star---TYC 1349-01421-1 indicates a distance of 131 pc. The molecular gases mainly concentrate on galactic plane, and for molecular clouds of which $b\sim 12 \degr$, the distance should be much less than 2.6 kpc; in addition, the second method shows a distance of $\sim$300 pc, which suggests that cloud is $<$ 1 kpc away, and the third method also indicates that the Gemini molecular cloud is a nearby cloud. On the other hand, the forth method also has a large uncertainty. Furthermore, the CO emissions suggest an averaged column density of Hydrogen of 0.7 $\times 10^{21}$ cm$^{-2}$ (see details in section 3.4.4), which indicates that extinction in the $J$ band ($A_{J}$) is 0.13 mag according to the relation of $N_{\mathrm{H}_{2}}=5.57\times 10^{21} A_J$ cm$^{-2}$ mag$^{-1}$ \citep{VMG2003}, and the $A_{K_{s}}\sim 0.1$ by using extinction curve of $A_{J}/A_{K_{s}}=0.282/0.114$ \citep{CCR1989}. The $A_{K_{s}}\sim 0.1$  suggests a distance of $\sim$ 120-170 pc, and also supports the cloud is far less than 1 kpc away, but the uncertainties of this relation are the scale coefficient and extinction curve, which are difficult to estimate errors here. And the $A_{K_{s}}\sim 0.1$ derived from the averaged column density of hydrogen may give an evidence that source J07084534+1610175 is most likely to be a background star. Combining those factors, we therefore adopt the distance of 400 pc, and denote it by $d_{ref}$. Because the distance estimated from the first method and averaged column density of Hydrogen is less than 200 pc, this distance may be overestimated. The units of any quantities depended on distance would multiply by an additional factor---$d/d_{ref}=d/(400\;\mathrm{pc})$, where $d$ is in unit of pc, which is the genuine distance to the Gemini molecular cloud.

\begin{figure}[!ht]
\centering

\includegraphics[width=8cm,height=8cm,angle=90]{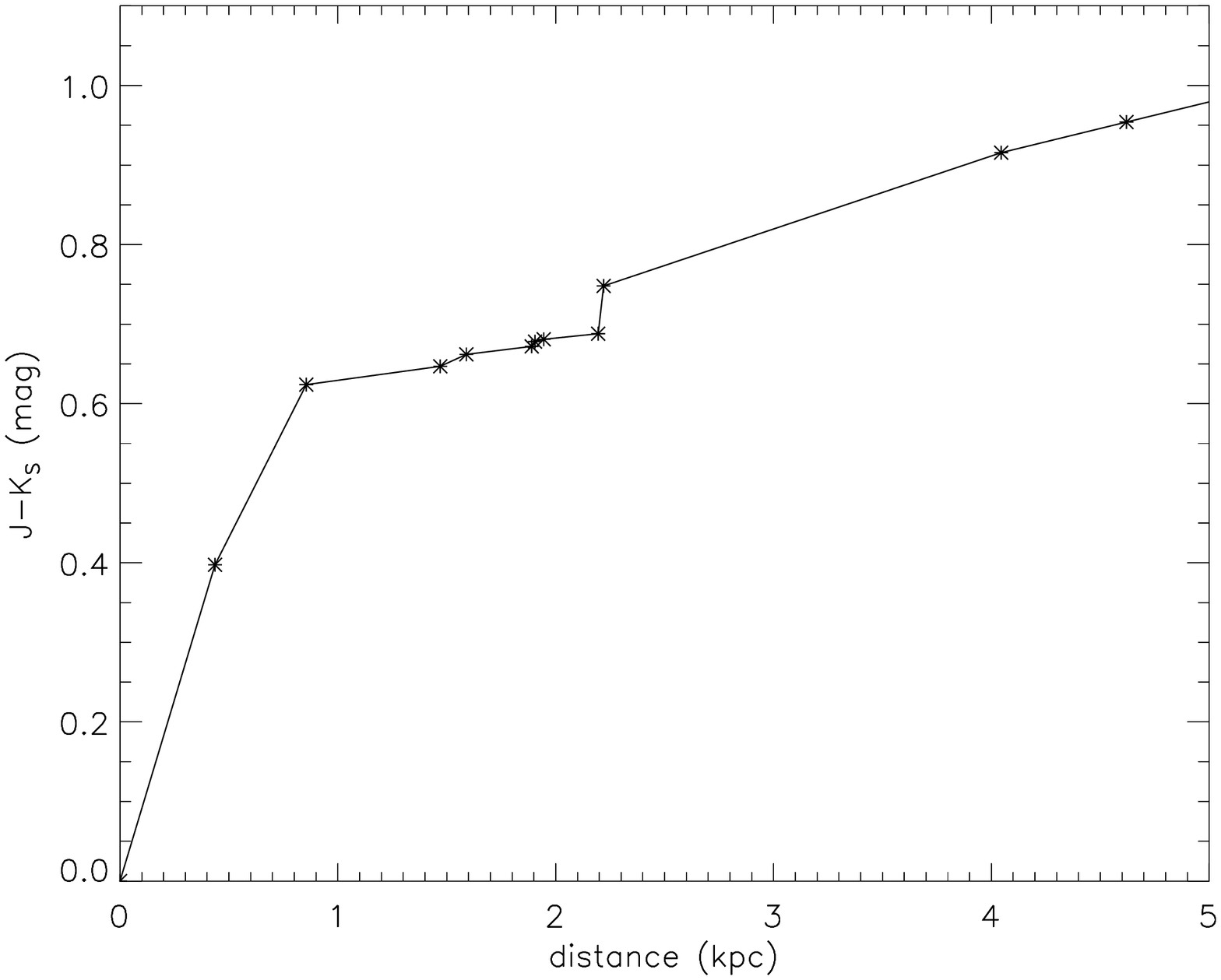}
\includegraphics[width=7.7cm,height=7.7cm,angle=0]{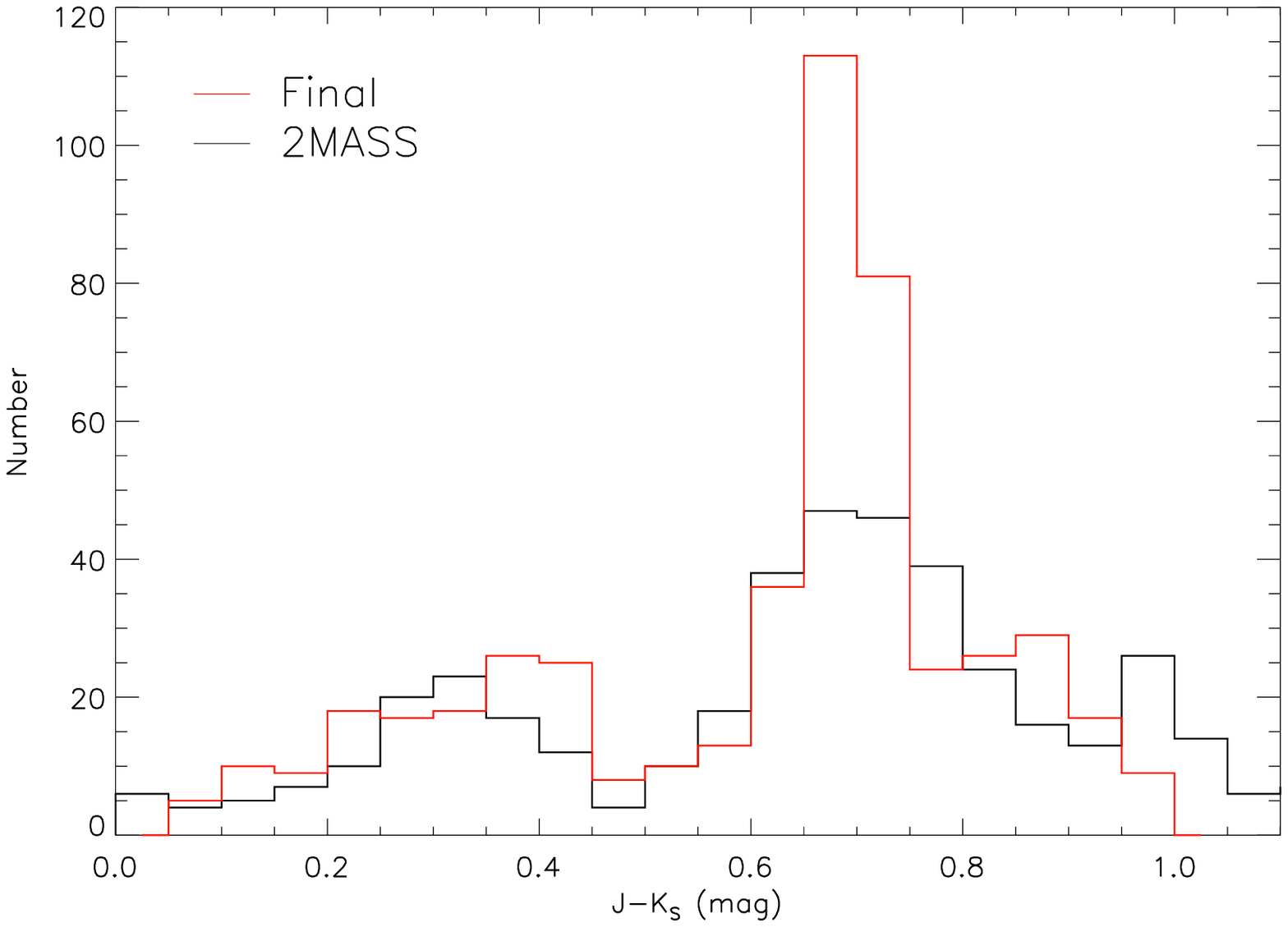}

\caption{Result of distance estimation by interstellar extinction distribution. left: $J-K_s$ .vs. distance, each asterisk represents a bin from our method. Right: corresponding $J-K_s$ histogram of the model and the observations. The black line represents the 2MASS observations and the red line is the result of our method.}
\end{figure}

\subsection{Core Identification}

The core candidate catalogue produced in this paper is created from the \xco data cube. We define cores as overdensities in a molecular cloud that may or may not contain cores from which single or multiple stars are born. These are also known as star-forming or starless cores, respectively \citep{PHCC, WBM2000}.

\footnotetext[1]{http://starlink.jach.hawaii.edu/starlink}

In order to decompose the Gemini molecular cloud into cores, we looked to automated core detection programs to deal with the obtained 3-dimensional FITS cube file of $^{13}$CO. In this paper, we used the CUPID (part of the STARLINK project software \footnotemark[1]) clump-finding algorithm CLUMPFIND \citep{WGB1994}, which has been widely used in the literature \citep{{M2007}, {B2010}}. Due to the intrinsic biases that are present in automated core detection routines \citep{SB2004, SCB2008}, we chose the core candidates carefully, while advising caution when performing blind cross-comparison of the properties of the core candidates measured here with those from other core detection algorithms. The CLUMPFIND algorithm first contours the data and searches for peaks in order to locate the core candidates, and then follows them down to lower intensities. To obtain as much of the emission as possible without contamination from noise, we set the parameters TLOW=2$\times$RMS and DELTAT=2$\times$RMS in $^{13}$CO, where TLOW determines the lowest level contour of a core candidate, and DELTAT represents the gap between contour levels that determines the lowest level at which to resolve merged core candidates \citep{WGB1994}. The parameters of each core candidate, including the position, velocity, size in the galactic longitude and galactic latitude directions, and one-dimensional velocity dispersions, are directly obtained using this process. Moreover, further steps were taken to ensure the detection of real cores. First, we excluded core candidates that had voxels touching the edge of the map area in order to avoid core candidates where the full extent of the emission may not have been recovered. Secondly, we excluded core candidates whose sizes were smaller than the beam resolution. Thirdly, we excluded core candidates that had ratios of two main-axes greater than 4. Finally, the morphology of the core candidates were checked by eye within the three-dimensional galactic longitude-galactic latitude-velocity space to identify core candidates with meaningful structures. Using these criteria, we obtain 36 core candidates (plotted in Figure 2 and catalogued in Table 2) in the $^{13}$CO maps.

\begin{table}[!ht]
 \begin{center}
 \textbf{Table 2}

 Properties of the 36 core candidates of \xco in the Gemini molecular cloud

 \end{center}
 \centering
 \begin{threeparttable}
 \scriptsize
 \begin{tabular} {lcccccccc} 
  \hline \hline
  Core name & $V_{lsr}$ & $\Delta v_{13}$ & $R_{13}$ & $T_{ex}$ & $\tau_{13}$ & $N_{\mathrm{H}_{2}}$ & $M$ & $M_V$ \\
   & (km s$^{-1}$) & (km s$^{-1}$) & ($(d/d_{def})$pc) & (K) & & ($\times 10^{21}$cm$^{-2}$) & ($(d/d_{def})^2$M$_{\sun}$) & ($(d/d_{def})$M$_{\sun}$) \\ 
   \hline
G198.81+12.29	&	-2.32 	&	0.40 	&	0.05 	&	7.7	&	0.77	&	1.4	&	0.3 	&	0.9 	\\
G198.87+12.22	&	1.66 	&	0.41 	&	0.11 	&	7.7	&	0.20	&	0.4	&	0.3 	&	2.4 	\\
G198.88+12.64	&	-1.83 	&	0.69 	&	0.13 	&	7.3	&	0.25	&	0.7	&	0.7 	&	7.3 	\\
G199.04+12.13	&	1.00 	&	0.33 	&	0.07 	&	7.4	&	0.59	&	0.8	&	0.3 	&	0.9 	\\
G199.35+11.72	&	1.66 	&	0.29 	&	0.05 	&	7.4	&	0.68	&	0.8	&	0.1 	&	0.5 	\\
G199.51+11.81	&	0.17 	&	0.77 	&	0.18 	&  10.7	&	0.09	&	0.5	&	1.0 	&	13.3 	\\
G199.55+11.87	&	1.49 	&	0.62 	&	0.13 	&	9.9	&	0.30	&	1.3	&	1.3 	&	5.8 	\\
G199.57+11.57	&	1.49 	&	0.63 	&	0.11 	&	6.9	&	0.31	&	0.7	&	0.7 	&	5.5 	\\
G199.72+11.80	&	0.17 	&	0.68 	&	0.09 	&	8.2	&	0.25	&	0.9	&	0.7 	&	5.8 	\\
G199.74+11.87	&	-0.17 	&	0.50 	&	0.15 	&	8.0	&	0.23	&	0.5	&	0.7 	&	4.4 	\\
G199.81+12.12	&	-0.83 	&	0.45 	&	0.07 	&	6.9	&	0.39	&	0.7	&	0.3 	&	2.0 	\\
G199.84+11.09	&	0.17 	&	0.50 	&	0.18 	&	9.1	&	0.26	&	0.8	&	1.7 	&	5.6 	\\
G199.95+11.77	&	3.98 	&	0.42 	&	0.09 	&	8.7	&	0.36	&	0.8	&	0.7 	&	2.2 	\\
G199.97+11.89	&	3.49 	&	0.63 	&	0.07 	&	9.5	&	0.27	&	1.1	&	0.3 	&	3.3 	\\
G199.98+12.07	&	3.32 	&	0.13 	&	0.07 	&	8.4	&	0.50	&	0.3	&	0.1 	&	0.2 	\\
G200.01+12.07	&	2.49 	&	0.48 	&	0.07 	&	8.4	&	0.30	&	0.7	&	0.3 	&	2.2 	\\
G200.03+11.90	&	2.99 	&	0.25 	&	0.09 	&	9.5	&	0.34	&	0.6	&	0.3 	&	0.7 	\\
G200.14+11.90	&	0.50 	&	1.60 	&	0.25 	&	8.2	&	0.16	&	1.3	&	5.3 	&	80.2 	\\
G200.15+12.22	&	-0.33 	&	0.85 	&	0.11 	&	6.8	&	0.35	&	1.1	&	1.0 	&	9.8 	\\
G200.18+12.37	&	-0.83 	&	0.81 	&	0.18 	&	7.7	&	0.25	&	0.9	&	2.3 	&	15.5 	\\
G200.20+11.57	&	0.00 	&	0.47 	&	0.13 	&	8.3	&	0.27	&	0.6	&	0.7 	&	3.3 	\\
G200.22+12.16	&	-1.66 	&	0.52 	&	0.11 	&	7.3	&	0.28	&	0.6	&	0.7 	&	4.0 	\\
G200.25+12.25	&	-0.33 	&	0.62 	&	0.09 	&	7.1	&	0.22	&	0.5	&	0.3 	&	4.5 	\\
G200.26+11.57	&	-0.66 	&	0.70 	&	0.11 	&	7.6	&	0.26	&	0.8	&	0.7 	&	6.9 	\\
G200.26+12.11	&	-1.16 	&	0.57 	&	0.18 	&	7.2	&	0.24	&	0.5	&	1.0 	&	7.1 	\\
G200.29+11.53	&	0.66 	&	0.63 	&	0.05 	&	7.6	&	0.28	&	0.8	&	0.3 	&	3.1 	\\
G200.29+11.04	&	-1.00 	&	0.35 	&	0.05 	&	8.4	&	0.39	&	0.7	&	0.1 	&	0.9 	\\
G200.31+12.21	&	-1.16 	&	0.37 	&	0.11 	&	6.7	&	0.36	&	0.5	&	0.3 	&	2.0 	\\
G200.37+11.49	&	-1.33 	&	0.43 	&	0.07 	&	7.5	&	0.37	&	0.7	&	0.3 	&	1.5 	\\
G200.42+11.77	&	-0.17 	&	0.58 	&	0.20 	&	8.2	&	0.23	&	0.7	&	2.0 	&	8.7 	\\
G200.46+11.47	&	1.00 	&	0.47 	&	0.36 	&	8.3	&	0.31	&	0.7	&	6.9 	&	10.4 	\\
G200.47+11.98	&	0.17 	&	0.49 	&	0.36 	&	7.4	&	0.32	&	0.7	&	5.6 	&	10.7 	\\
G200.58+11.77	&	-0.33 	&	0.17 	&	0.07 	&	7.6	&	0.54	&	0.4	&	0.1 	&	0.2 	\\
G200.61+11.76	&	-1.00 	&	0.41 	&	0.09 	&	7.6	&	0.44	&	0.8	&	0.3 	&	1.8 	\\
G200.63+11.82	&	0.50 	&	0.60 	&	0.09 	&	7.6	&	0.33	&	0.9	&	0.3 	&	4.2 	\\
G200.69+11.77	&	-0.17 	&	0.54 	&	0.09 	&	7.0	&	0.31	&	0.6	&	0.3 	&	3.1 	\\
mean	        &	0.32 	&	0.54 	&	0.12 	&	7.9	&	0.33	&	0.7	&	1.1 	&	6.7 	\\

\hline

\end{tabular}
 \end{threeparttable}

 \end{table}

Properties of the 36 core candidates of \xco are summarized in Table 2. The columns represent, respectively, the catalogue core (candidate) name, $V_{lsr}$, the  line width ($\Delta v_{13}$, i.e., the velocity component along the line of sight, which is fitted by Gaussian profile), its radius ($R_{13}$), its excitation temperature ($T_{ex}$), the optical depth ($\tau_{13}$), the column density of hydrogen molecular ($N_{\mathrm{H}_{2}}$), its LTE mass ($M$) and its virial mass ($M_V$). Core candidate names also convey their location, for instance: G199.04+12.13 represents the galactic longitude 199.04$\degr$ and galactic latitude 12.13$\degr$. The last row in Table 2 is the mean value of each physical quantity.

\subsection{Physical Properties}

\subsubsection{Radius}

The core candidate radii are derived from the geometric mean of the core candidate sizes in two directions, and we have made the size of the beam deconvolved from the radius measurement. Figure 7 shows the distributions of the core candidate radii in $^{13}$CO. The mean core candidate radius in \xco is 0.12 $(d/d_{ref})$ pc, and the median core candidate radius of \xco is 0.10 $(d/d_{ref})$ pc, which is slightly less than the mean value.

\begin{figure}[!ht]
\centering

\includegraphics[width=9.1cm,height=9cm,angle=0]{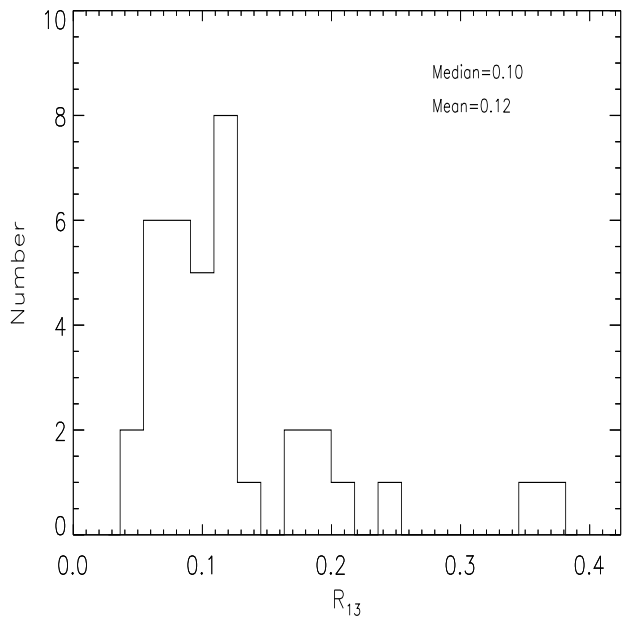}

\caption{Distribution of core candidate radii in $^{13}$CO (in unit of $(d/d_{ref})$ pc). The mean and median values (in unit of $(d/d_{ref})$ pc) are marked on the upper-right corner of this plot.}
\end{figure}

\subsubsection{Excitation Temperature}

To obtain the excitation temperature of each core candidate, three assumptions are required. First, the core candidates are in local thermodynamic equilibrium (LTE); secondly, the emission line used in this determination is optically thick; and thirdly, the emission fills the beam. Assuming $T_{ex}(^{13}\mathrm{CO})=T_{ex}(^{12}\mathrm{CO})=T_{ex}$ \citep{KM1986}, we calculated the  excitation temperatures of the \co emission using the following:

\begin{equation}
\centering
T_{ex}=h\nu\left(k \ln  \left(1+ \left(\frac{kT^{\ast}_R(^{12}\rm{CO})}{h\nu}+\frac{1}{\exp\left(h\nu/kT_{bg}\right)-1}\right)^{-1}\right)\right)^{-1},
\end{equation}

\noindent where $T_{ex}$ is in unit of K, $h\nu$ is the energy of a single photon emitted during the transition of $^{12}$CO(J=1-0), $k$ is the Boltzmann constant, $T^{\ast}_{R}(^{12}\rm{CO})$ is the peak main beam temperature in unit of K (for \co emission), and $T_{bg}$ is 2.7 K, i.e., the temperature of cosmic microwave background radiation.

Excitation temperatures of the 36 core candidates catalogued in Table 2 range from 6.7 to 10.7 K, with a mean temperature of 7.9 K. We note that the excitation temperatures may be underestimated when we consider the effects of beam dilution and CO self-absorption on the obtained observations.

\subsubsection{Opacity}

The opacity of \xco is given by  \citep{KOY1998}:

\begin{equation}
\centering
{{\tau_{13} \approx
-\ln \left(1-\frac{T^{\ast}_{R}(^{13}\rm{CO})}{5.29\left(1/\left(\exp\left(5.29/T_{ex}\right)-1\right)-0.164\right)}\right)},}
\end{equation}

\noindent where $T_{ex}$ and $T^{\ast}_{R}(^{13}\rm{CO})$ (in unit of K) are given in Section 3.4.2. This formula indicates that uncertainties in the excitation temperature will directly affect the opacity. The opacities of the 36 core candidates are catalogued in Table 2. We have calculated the ratio of $T^{\ast}_{R}(^{12}\rm{CO})/T^{\ast}_{R}(^{13}\rm{CO})$ in the position of peak emission of those \xco core candidates, and this ratios range from 3.1 to 7.2, with mean ratio of 4.2, and we then find that the \co is indeed optically thick with $\tau_{12}> 1$, and the \co opacity of 58.4\% core candidates are larger than 5.

\subsubsection{Column Density}

The column density of \xco $N(^{13}\mathrm{CO})$ is given by \citep{KOY1998}

\begin{equation}
\centering
N(^{13}\mathrm{CO})=2.42\times10^{14}\times\frac{\tau_{13}T_{ex}\Delta v_{13}}{1-\exp\left(-5.29/T_{ex}\right)},
\end{equation}

\noindent where the units of $N(^{13}\mathrm{CO})$ are cm$^{-2}$, $T_{ex}$ is the excitation temperature of the J=1-0 transition of \xco in unit of K. The line width $\Delta v_{13}$ is in unit of km s$^{-1}$. The value of $T_{ex}$, $\tau_{13}$ and $\Delta v_{13}$ are catalogued in Table 2.

Column densities in terms of H$_{2}$ ($N_{\mathrm{H}_{2}}$) as opposed to \xco are determined using the abundance ratio between  H$_{2}$ and $^{13}$CO \citep{W1999, MBM1985}, and the value adopted here is 1.2$\times$10$^{6}$. $N_{\mathrm{H}_{2}}$ of the 36 core candidates catalogued in Table 2 range from 3.4$\times$10$^{20}$ to 1.4$\times$10$^{21}$ cm$^{-2}$ with a mean value of 7.5$\times$10$^{20}$ cm$^{-2}$ (see table 2).  Figure 8 shows the distribution of $N_{\mathrm{H}_{2}}$. Systematic errors in the column densities derived here may arise from both their opacity and their abundance ratio.

\begin{figure}[!ht]
\centering

\includegraphics[width=9cm,height=9cm,angle=0]{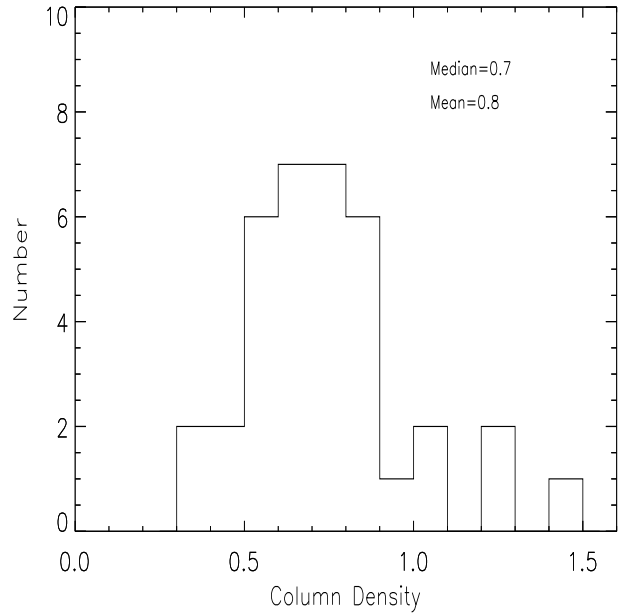}

\caption{Distribution of column densities (in unit of $\times10^{21}$ cm$^{-2}$) of molecular hydrogen. The mean and median values (in unit of $\times10^{21}$ cm$^{-2}$) are marked in the upper-right corner.}
\end{figure}

Surface densities in unit of g cm$^{-2}$ are calculated by multiplying the column densities by the mass of molecular hydrogen.  Within the observed Gemini molecular hydrogen, the surface densities range from 1.5$\times$10$^{-3}$ to 6.4$\times$10$^{-3}$ g cm$^{-2}$, with mean value of 3.4$\times$10$^{-3}$ g cm$^{-2}$.

\subsubsection{Mass}

Once the density and size of a core candidate have been determined, the LTE mass ($M$, in unit of $(d/d_{ref})^2$ M$_{\sun}$) of the core candidate reads

\begin{equation}
\centering
M=AN_{^{13}\mathrm{CO}}\frac{[\mathrm{H}_{2}]}{[^{13}\mathrm{CO}]}\mu_{\mathrm{H}_{2}}m_{\mathrm{H}},
\end{equation}

\noindent where $A=\pi(R_{13})^{2}$ is the area of the core candidate in unit of cm$^{2}$, where $R_{13}$ (in unit of
$(d/d_{ref})$ pc) is the \xco core candidate radius which is given in Table 2 and section 3.4.1. $N_{^{13}\mathrm{CO}}$ is the column density of \xco in unit of cm$^{-2}$, as given in section 3.4.4., and the abundance ratio is again assumed as [H$_{2}$]/[$^{13}$CO]$\sim1.2\times10^{6}$. $\mu_{\mathrm{H}_{2}}$ is the mean molecular weight of gas per $\mathrm{H}_{2}$ molecular \citep[$\mu_{\mathrm{H}_{2}}=2.72$ that includes hydrogen, helium and the isotopologues of carbon monoxide; e.g.,][]{B2010} and $m_{\mathrm{H}}$ is the mass of a single hydrogen atom.

The core candidate masses estimated here, which are based on the assumption of LTE, range from 0.1 to 6.9 $(d/d_{ref})^2$ M$_{\sun}$ with mean mass of 1.1 $(d/d_{ref})^2$ M$_{\sun}$ (see Table 2). The mass error comes from the distance error and its abundance ratio. Because some CO emission is absorbed by grains, these estimates can be considered as lower limits, and an estimate of the intervening dust mass needed to be investigate to determine how much it will affect our observations. The total mass of these 36 core candidates is 38.3 $(d/d_{ref})^2$ M$_{\sun}$.

To obtain the virial mass $M_V$, the radius and line width are required, and a core density profile of $\rho\propto r^{-2}$ is assumed. The virial mass $M_V$ reads:

\begin{equation}
\centering
M_V=126R_{13}(\Delta v)^{2},
\end{equation}

\noindent where $M_V$ is the virial mass in unit of $(d/d_{def})$ M$_{\sun}$, $R_{13}$ is the radius in $(d/d_{ref})$ pc, and $\Delta v$ is the line width in km s$^{-1}$ \citep{MRW1988}. The uncertainties in the virial masses come from the measurement errors of line widths and $R_{13}$.

The virial masses derived for \xco are about 7.0 $d_{ref}/d$ times larger than the LTE masses, and have a range of 0.2 to 80.2
$(d/d_{def})$ M$_{\sun}$. The a mean virial mass is 6.7 $(d/d_{def})$ M$_{\sun}$ (see Table 2).

\subsubsection{$V_{lsr}$ and Line Width}

The core candidate line widths increase with increasing LTE masses, however the excitation temperatures do not appear to have this tendency. This implies that thermal motion is not the exclusive broadening mechanism due to $\sigma_{Thermal}\approx (T/A)^{1/2}$ km s$^{-1}\propto T^{1/2}$ in a coarse sense, where $\sigma_{Thermal}$ is the thermal line width (line width broadened via pure thermal motion), $T$ is the kinetic temperature in unit of K, which $\gtrsim T_{ex}$, and $A$ is the atomic mass number.

To determine the dominant broadening mechanisms, we calculate thermal line widths of tracer species in unit of km s$^{-1}$ using $\sigma_{Thermal}=\sqrt{kT/m}/1000$, where $k$ is the Boltzmann constant, $m$ is the mean molecular mass in unit of kg and $T$ is the kinetic temperature, which $\gtrsim T_{ex}$, in unit of K. The non-thermal line width $\sigma_{Non-Thermal}$ and thermal line widths of gas as a whole $\sigma_{Thermal,g}$ read:
\begin{figure}[!ht]
\centering

\includegraphics[width=8cm,height=8cm,angle=0]{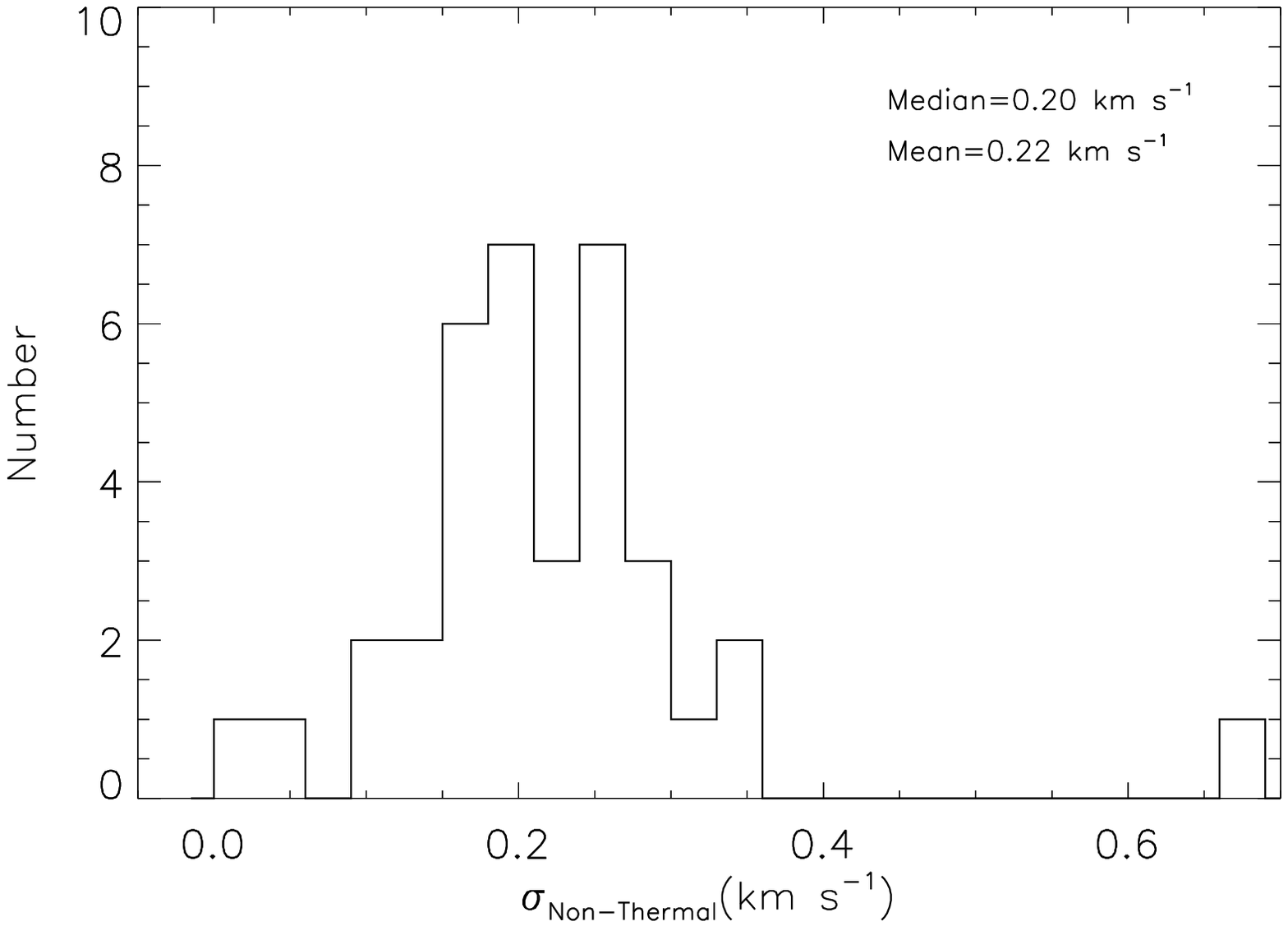}

\includegraphics[width=8cm,height=16cm,angle=90]{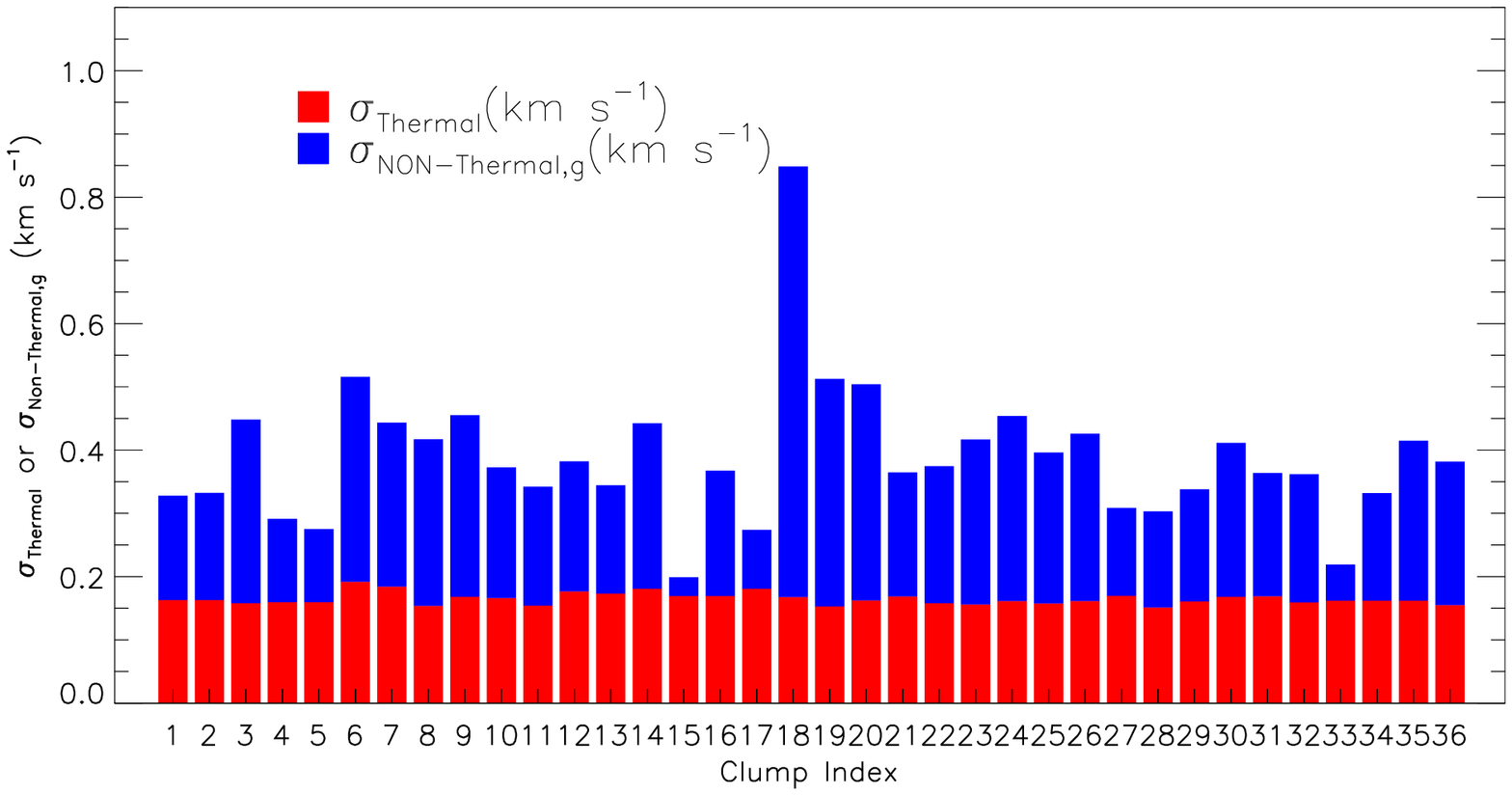}

\caption{Top: histogram of non-thermal line widths. The mean value and median value are marked in the right-upper corner of the histogram. Bottom: histogram of non-thermal (blue) line width and thermal (red) line width of gas as a whole. We numbered the \xco core candidates in ascending order of galactic longitude, and marked them on the x-label for a brief view.}
\end{figure}
\begin{equation}
\centering
\sigma_{Non-Thermal}=\sqrt{\sigma_{1D}^{2}-\sigma_{Thermal}^2}=\sqrt{\Delta v_{13}^{2}/(8 \ln 2)-\sigma_{Thermal}^2},
\end{equation}

\begin{equation}
\centering
\sigma_{Thermal,g}=\frac{1}{1000}\sqrt{\frac{kT}{\mu m_{\mathrm{H}}}},
\end{equation}

\noindent  where $\sigma_{1D}$ is the one-dimensional velocity dispersion in unit of km s$^{-1}$, and $\Delta v_{13}$ is the line width in unit of km s$^{-1}$ \citep[see e.g.,][]{M1983}; and $k$ is the Boltzmann constant, $\mu =2.4$, $m_{\mathrm{H}}$ is the mass of a single hydrogen atom in unit of kg and $T$ is the kinetic temperature, which $\gtrsim T_{ex}$, in unit of K. $\sigma_{Thermal}$ ranges from 0.04 to 0.06 km s$^{-1}$, $\sigma_{Thermal,g}$ ranges from 0.15 to 0.19 km s$^{-1}$and the distribution of $\sigma_{Non-Thermal}$ is plotted in the top histogram of Figure 9. Because $T \gtrsim T_{ex}$, the $\sigma_{Thermal}$ and $\sigma_{Thermal,g}$ is underestimated (i.e., a lower limit) and $\sigma_{Non-thermal}$ is therefore overestimated (i.e., a upper limit).

The mean ratio of $\sigma_{Non-Thermal}$ to $\sigma_{Thermal,g}$ is 1.35 (see the bottom histogram in Figure 9), which implies that the non-thermal broadening mechanism plays a dominant role in the core candidates; or to be more details, 19\% (i.e. 7/36) of those are subsonic core candidates, and the rest are supersonic core candidates, in which the non-thermal broadening mechanism plays a dominant role.

The core candidates of high $V_{lsr}$ (greater than 2 km s$^{-1}$) and low $V_{lsr}$ (lower than -1.5 km s$^{-1}$) in \xco belong to two different regions (i.e., map 1 and map 2 respectively) which are plotted in the channel maps seen in Figure 4.

The big change of morphology in the channel map in Figure 4, imply that the velocity gradients mentioned in section 3.1 (i.e., 0.3 km s$^{-1}$ pc$^{-1}$) and the large dispersion of $V_{lsr}$ of the \xco core candidates (i.e., 6.3 km s$^{-1}$) in the Gemini molecular cloud may be more strongly affected by stochastic processes, such as collisions and chaotic magnetic fields, rather than ordered motions such as rotation.

\section{Discussion}

In this section, we will discuss the statistical properties and star-formation processes occurring in the Gemini molecular cloud. There is no sharp demarcation between these two aspects. In fact, statistical properties can be regarded as tools in research of star-formation.

\subsection{Core Candidate Mass Function}

The core mass function (CMF) describes the relative frequency of cores with differing masses. CMF is commonly fitted with a basic power law $dN/dM\propto M^{-\alpha}$, where $N$ is the core candidates in each bin, $M$ is the mass and $\alpha$ is the corresponding power index. The shape of the CMF has been seen to resemble the stellar initial mass function (IMF) that describes the relative frequency of stars with differing masses over a large range of environments \citep{S1955, SNW2008}. Understanding the relationship between the CMF and the IMF can help constrain star formation models \citep{BB2005, RW2006}.

To obtain the parameter index of the CMF, we fixed the bin widths and counted the number of core candidates per bin. The value of abscissa is the mean value in each bin, and two kinds of CMF derived from \xco are given by:
\begin{equation}
\centering
\frac{dN}{dM}\propto M^{-\alpha_{1}},
\end{equation}

\begin{equation}
\centering
\frac{dN}{dM_{V}} \propto M^{-\alpha_{2}}_{V},
\end{equation}

\begin{figure}[!ht]
\centering

\includegraphics[width=7.1cm,height=8.1cm,angle=90]{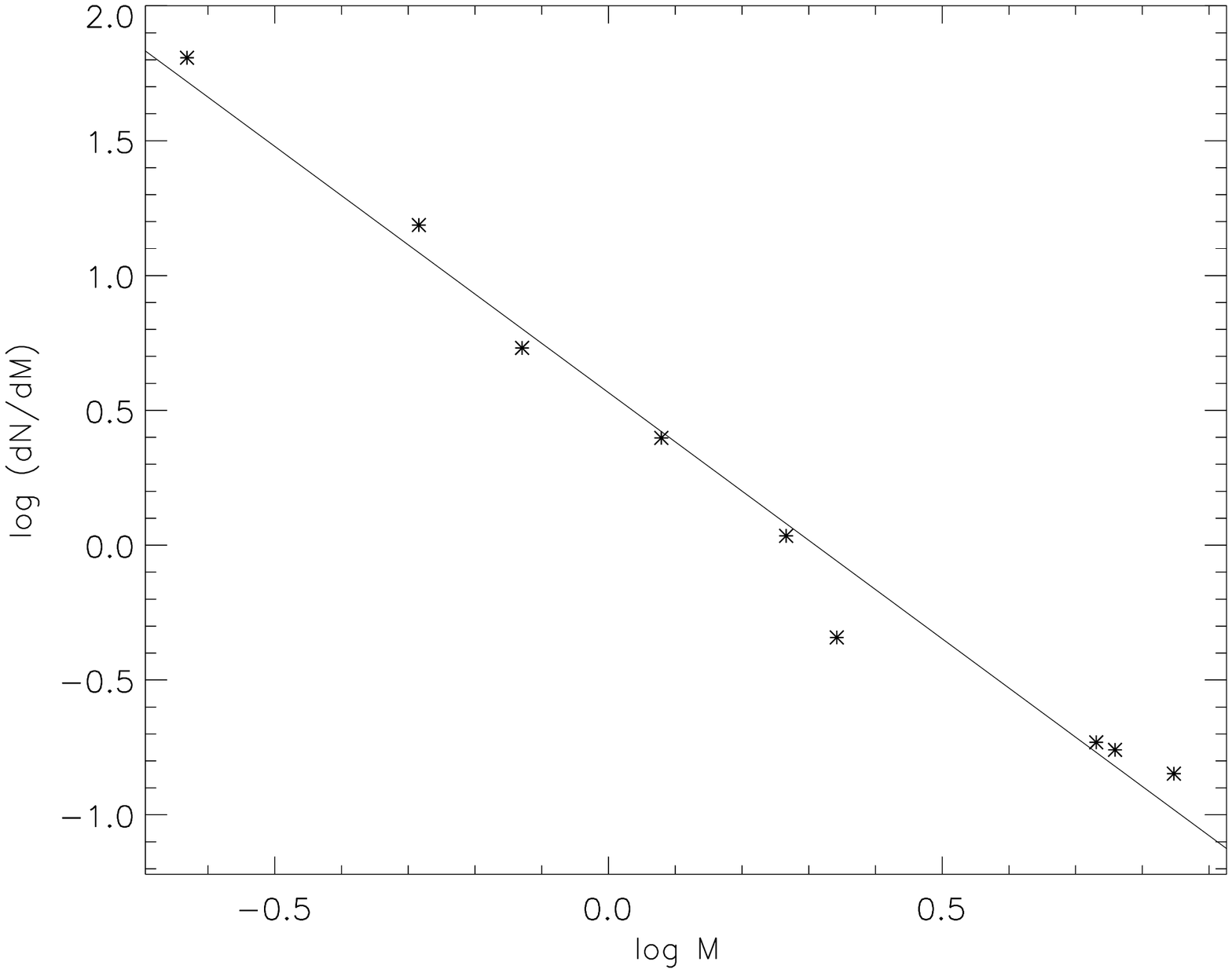}
\includegraphics[width=7.1cm,height=8.1cm,angle=90]{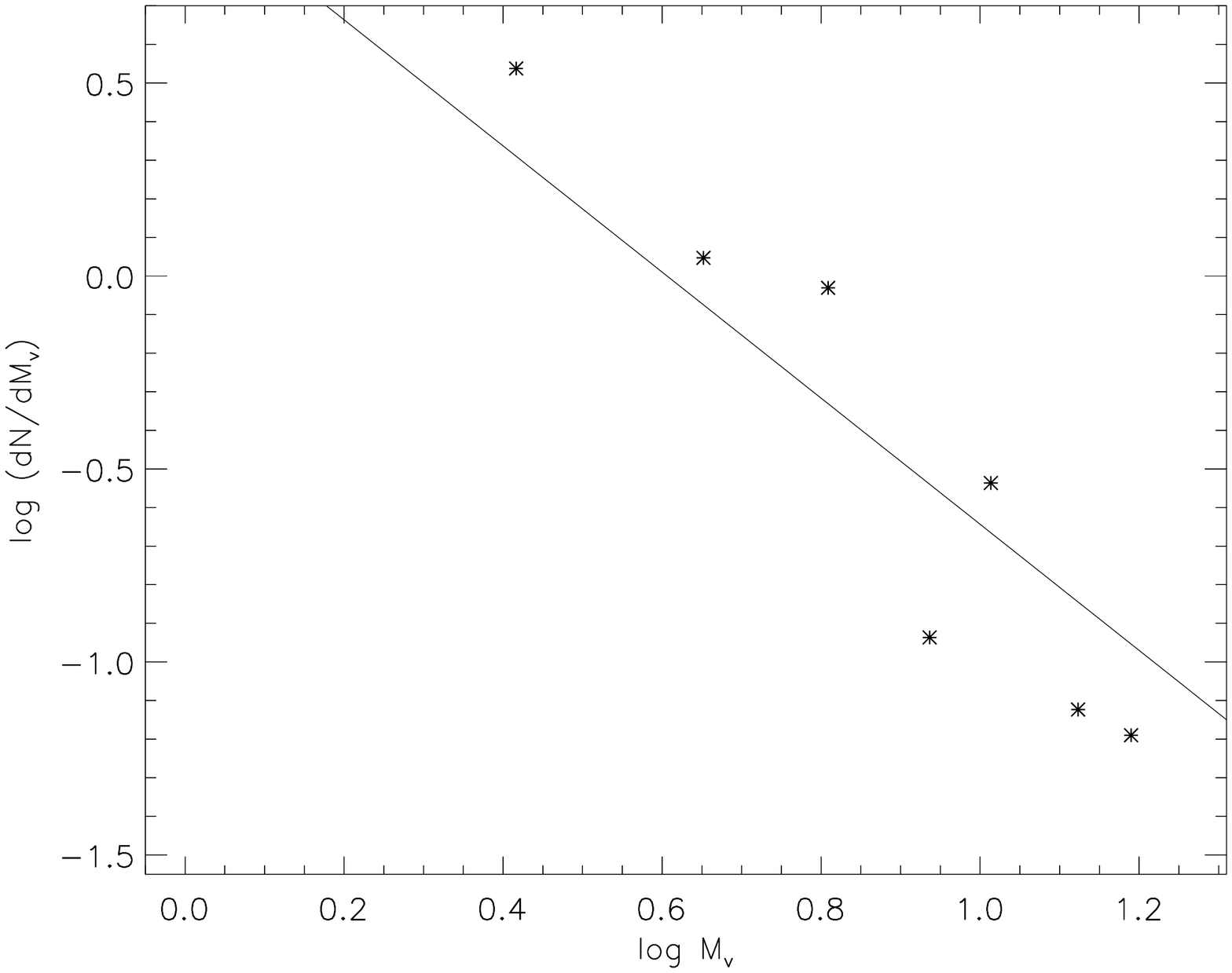}

\caption{CMF (where mass includes $M$ in the units of $(d/d_{def})^2$ M$_{\sun}$ and $M_{V}$ in unit of
 $(d/d_{def})$ M$_{\sun}$). The black star-like points are logarithmic values of the number of core candidates per unit mass against $\log M$ (left) and $\log M_{V}$ (right). $\alpha_{1}$ and $\alpha_{2}$ are 1.83$\pm$0.12 and 1.63$\pm$0.12,  respectively. $\alpha_{1}$ and $\alpha_{2}$  correspond to power the index of CMF of LTE mass and viral mass as derived from $^{13}$CO, respectively.}
\end{figure}

\noindent where $M$ and $M_{V}$ are the LTE masses (in unit of $(d/d_{ref})^2$ M$_{\sun}$) and virial masses (in unit of $(d/d_{ref})$ M$_{\sun}$) derived from $^{13}$CO, respectively. $N$ is the number of core candidates in each bin, while $\alpha_{1}$ and $\alpha_{2}$ are the corresponding power indices. The power law correlations of CMF are: $dN/dM=(0.57\pm 0.04)M^{-1.83\pm0.07}$(left), and $dN/dM_V=(0.99\pm 0.11)M_V^{-1.63\pm0.12}$(right), see Figure 10. Note, the significance of the coefficients of $0.57\pm 0.04$ and $0.99\pm 0.11$ is very limited for the uncertainty of distance. Our power index (1.83 for LTE mass and 1.63 for virial mass derived from \xco, see Figure 10) is between the IMF of 2.35 \citep{S1955} and CMF of 1.6 -1.8 presented by \citet{KSRC1998} and \citet{B2010}.

The similarity between the CMF and IMF power indices can be simply be explained by a constant star formation efficiency that is unrelated to the mass and self-similar cloud structure, and based on a scenario of one-to-one transformation from cores to stars \citep{LMR2008}. In addition, a simulation by \citet{SW2008} suggested that the obtained IMF is very similar to the input CMF even when different fragmentation modes are considered. However, the full physical meaning of the relationship between the CMF and the IMF remains unclear due to a number of complications, including: (a) completeness limitations, (b) time-scales, (c) physical size scales, (d) distances, etc. \citep{CR2010, RWPS2010}.

\subsection{Star Formation}

Figure 11 shows the relationship between the virial mass $M_{V}$ and LTE mass $M$. The virial parameter $\alpha_{V}$, defined as $M_{V}/M$, describes the competition of internal supporting energy against the gravitational energy. Virial parameter is inversely proportional to distance, and its typical value is 7.0 $(d_{ref}/d)$, and all virial parameters for the 36 core candidates in \xco are larger than 1.9 $(d_{def}/d)$. The massive core candidates tend to have lower virial parameters. Furthermore, a linear fitting between LTE mass ($M$ in unit of $(d/d_{ref})^2$ M$_{\sun}$) and virial mass ($M_{V}$ in unit of $(d/d_{ref})$ M$_{\sun}$) may be a good indicator of the typical virial parameter: the first-order coefficient is 4.84 $(d_{def}/d)$, as shown in Figure 11, which is slightly less than the typical virial parameter value of 7.0 $(d_{ref}/d)$. However, the correlation coefficient of this linear correlation is just 0.60.

The mass relationship can also be fitted with a power-law of $M_{V}\propto M^{\gamma}$ (where $M$ in unit of
$(d/d_{ref})^2$ M$_{\sun}$ and $M_V$ in unit of $(d/d_{ref})$ M$_{\sun}$), and the result is $M_V=(0.76\pm 0.15)M^{0.97\pm 0.15}$, $c.c.=0.85$, where $c.c.$ is the correlation coefficient, and the significance of the coefficient of $0.76\pm 0.15$ is very limited for the uncertainty of distance.
\begin{figure}[!ht]
\centering

\includegraphics[width=9cm,height=10cm,angle=90]{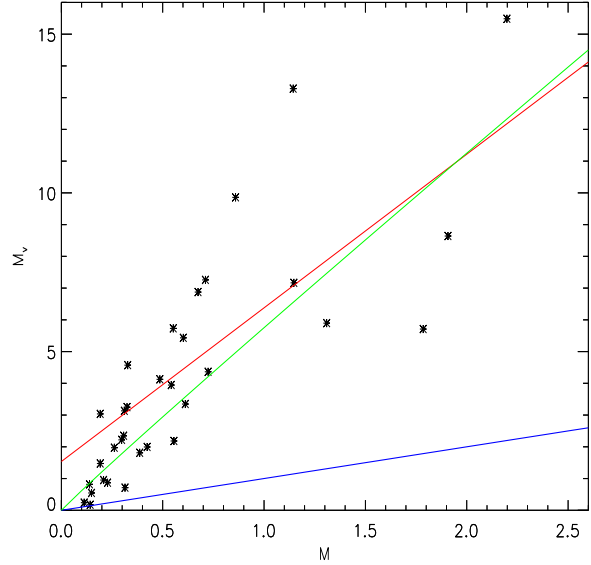}

\caption{Virial mass $M_{V}$ (in unit of $(d/d_{ref})$ M$_{\sun}$) .vs. LTE mass $M$ (in unit of $(d/d_{ref})^2$ M$_{\sun}$) relation of the core candidates. The blue slanted line: the slope is 1.0, denotes the minimum value of virial parameter. The red slanted line: the slop is 4.84, and the correlation coefficient is 0.60, presents the linear correlation among virial mass and LTE mass. The green curve: shows the power-law correlation among virial mass and LTE mass, and the result is $M_V=(0.76\pm 0.15)M^{0.97\pm 0.15}$, $c.c.=0.85$, where $c.c.$ is the correlation coefficient.}
\end{figure}

The power index value of 0.97 that we obtained is larger than the value of 0.67 in Orion B (which are gravitationally bound) reported by \citet{IK2009} and 0.61 in Planck cold clumps (which are gravitationally bound) presented by \citet{LWZ2012}, and is significantly higher than the index of pressure-confined clumps ($\alpha_{V}\propto M^{-2/3}$) given by \citet{BM1992}. While the virial parameters are larger than 1.9, which indicates that the core candidates may be unbound.

Figure 11 also shows that the virial mass changes more dramatically in the lower end of the LTE mass than in the upper end of the LTE mass. Additionally, the ratio between the virial and LTE masses decreases as LTE mass increases.

However, the discussion of whether these core candidates are pressure confined or not may be not necessary if such structures do not need to be long lived. We then calculate a time-scale ($\tau_{l}$) used by \citet{L1981}:

\begin{equation}
\centering
 \tau_{l}\sim 4.6\times 10 ^6\frac{R_{13}}{2\sqrt{2 \ln 2}\sigma_{t}}\frac{d}{d_{ref}}\; yr,
\end{equation}

\noindent where, $R_{13}$ is in unit of pc, and $\sigma_{t}=\sqrt{(\sigma_{Non-thermal})^2+(\sigma_{Thermal,g})^2}$ is the overall velocity dispersion of the mean particle in the cloud, where $\sigma_{t}$, $\sigma_{Non-thermal}$ and $\sigma_{Thermal,g}$ are in unit of km s$^{-1}$. The typical value of $\tau_{l}$ is about $1.7\times10^6$ yr. And thus, the Gemini molecular cloud may be transient structure.

Cores that do not have a known infrared associations are starless cores or pre-protostellar cores \citep{BM1989, WSHA1994}. Those cores represent a stage earlier than protostellar. We have checked the IRAS point source  catalogue in this region and found that no \xco core candidates are associated with IRAS point sources. We also have checked the WISE 22 $\mathrm{\mu}$m data, and found that only one core candidate (i.e. Core candidate G200.15+12.22) is associated with WISE source J071257.49+164743.5, which has instrumental profile-fit photometry magnitude of 9.0 mag in the band of 22 $\mathrm{\mu}$m. However, the columns of w4sigmpro = ``null'' and w4chi = 0.9, which indicates that this source is not measurable in the band of 22 $\mathrm{\mu}$m, where the first two letters of ``w4'' in each column mean band 4 with wavelength of 22 $\mathrm{\mu}$m. We then conclude that all 36 \xco core candidates may be starless core candidates. Nevertheless, to determine whether these core candidates are in the process of forming stars, or already formed stars, or even other situations, further study is necessary.

\section{Summary}

We presented PMODLH mapping observations for an area of 4 deg$^{2}$ toward the Gemini molecular cloud centered at \emph{l} = 200$\degr$ and \emph{b} = 12$\degr$ in $^{12}$CO, \xco and C$^{18}$O lines. The main results are summarized as follows:

1. We identified 36 core candidates in $^{13}$CO. Derived from $^{13}$CO, we yield typical radii, column densities and LTE masses of 0.12 $(d/d_{ref})$ pc, 7.5$\times 10^{20}$ cm$^{-2}$ and 1.1 $(d/d_{def})^2$ M$_{\sun}$ respectively. We also found the mean excitation temperature of 7.9 K as derived from $^{12}$CO. The total LTE mass derived from these 36 core candidates in \xco is 38.3 $(d/d_{ref})^2$ M$_{\sun}$.

2. Non-thermal broadening mechanism plays a dominant role in 81\% (29/36) of those core candidates.

3. Velocity gradients (i.e., 0.3 km s$^{-1}$ pc$^{-1}$) and large dispersions in $V_{lsr}$ of the \xco core candidates (i.e., 6.3 km s$^{-1}$) in the cloud may be more largely affected by stochastic processes, such as collisions, and chaotic magnetic fields rather than ordered motions such as rotation.

4. Two kinds of CMF are present in this paper, namely the CMF of LTE masses, and the virial masses derived from $^{13}$CO. Their power indices are $\alpha_{1}=1.83$ and $\alpha_{2}=1.63$, respectively. $\alpha_{1}$ and $\alpha_{2}$ are all lower than the power index (2.35) of the IMF. Moreover, the CMF of LTE mass is steeper than the CMF of the virial mass derived from $^{13}$CO.

5. \xco core candidates in the Gemini molecular cloud are more likely unbound, and massive core candidates have lower virial parameters. All 36 core candidates are potential starless core candidates, and they also may be transient structure.

\acknowledgments
We would like to thank the referee for many useful comments that have improved the paper. We also acknowledge Xiaolong-Wang, Chong Li and ShaoBo Zhang for their valuable help. This work is supported by the National Natural Science Foundation of China (grant Nos. 11133008 and 11233007), the Strategic Priority Research Program of the Chinese Academy of Sciences (grant No. XDB09010300), and the Key Laboratory for Radio Astronomy.

\end{document}